\documentclass[iop,apj]{emulateapj}
\usepackage{amsmath,amssymb,amstext}

\usepackage[breaklinks,colorlinks,citecolor=blue,linkcolor=magenta]{hyperref} 

\usepackage[all]{hypcap} 

\usepackage{aas_macros}
\usepackage{natbib}
\newcommand{\fesc}{f_{\mathrm{esc}}}

\shorttitle{Escape fractions with machine learning}
\shortauthors{Jensen et al.}

\begin{document}

\title{A machine-learning approach to measuring the escape of ionizing radiation from galaxies in the reionization epoch}
\author{Hannes Jensen$^{1}$, Erik Zackrisson$^{1}$, Kristiaan Pelckmans$^2$, Christian Binggeli$^{1}$, Kristiina Ausmees$^2$, Ulrika Lundholm$^2$}
\affil{$^1$Department of Physics and Astronomy, Uppsala University, Box 515, SE-751 20 Uppsala, Sweden \\
$^2$Department of Information Technology, Division of Systems and Control (Syscon), \\ Uppsala University, Box 337, SE-751 05 Uppsala, Sweden}
\email{hannes.jensen@physics.uu.se}

\begin{abstract}
Recent observations of galaxies at $z \gtrsim 7$, along with the low value of the electron scattering optical depth measured by the Planck mission, make galaxies plausible as dominant sources of ionizing photons during the epoch of reionization. However, scenarios of galaxy-driven reionization hinge on the assumption that the average escape fraction of ionizing photons is significantly higher for galaxies in the reionization epoch than in the local Universe. The NIRSpec instrument on the James Webb Space Telescope (JWST) will enable spectroscopic observations of large samples of reionization-epoch galaxies. While the leakage of ionizing photons will not be directly measurable from these spectra, the leakage is predicted to have an indirect effect on the spectral slope and the strength of nebular emission lines in the rest-frame ultraviolet and optical. Here, we apply a machine learning technique known as lasso regression on mock JWST/NIRSpec observations of simulated $z=7$ galaxies in order to obtain a model that can predict the escape fraction from JWST/NIRSpec data. Barring systematic biases in the simulated spectra, our method is able to retrieve the escape fraction with a mean absolute error of $\Delta \fesc \approx 0.12$ for spectra with $S/N\approx 5$ at a rest-frame wavelength of 1500~\AA~for our fiducial simulation. This prediction accuracy represents a significant improvement over previous similar approaches.
\end{abstract}
\maketitle

\section{Introduction}
Current constraints on the reionization of the Universe are consistent with scenarios in which star-forming galaxies at $z\gtrsim 6$ provide the majority of hydrogen-ionizing (Lyman continuum, LyC) photons to bring about this phase transition \citep[e.g.][]{Robertson15,Mitra15}. However, the notion of galaxy-dominated reionization relies on the assumption that a non-negligible fraction of the LyC photons produced by hot, young stars within these objects can avoid absorption by gas and dust in the interstellar medium and  make it into the intergalactic medium (IGM). At the current time, it remains unclear if LyC leakage from $z\gtrsim 6$ galaxies is really taking place at the level required to make this work, or whether alternative sources such as quasars may be required \citep[e.g.][]{Madau15}.
	
The production rate of LyC photons entering the IGM from the galaxy population at redshift $z$ can be written as:
\begin{equation}
\dot{N}_\mathrm{ion}(z) = \fesc (z)\xi_\mathrm{ion}(z) \rho_\mathrm{UV}(z),
\end{equation}
where $\dot{N}_\mathrm{ion}$ is the comoving LyC photon production rate density (a.k.a. the ionizing emissivity; photons~s$^{-1}$~Mpc$^{-3}$), $\fesc$ is the LyC escape fraction and $\xi_\mathrm{ion}$ is the production efficiency of Lyman continuum photons (photons~erg$^{-1}$~Hz; the rate of LyC photon production relative to the luminosity in the rest-frame, non-ionizing ultraviolet continuum, usually at 1500 \AA) and $\rho_\mathrm{UV}$ is the luminosity density (erg~Hz$^{-1}$~Mpc$^{-3}$) of the galaxy population in the rest-frame ultraviolet (UV) continuum. The UV luminosity density $\rho_\mathrm{UV}$ can be constrained from observations of galaxy luminosity functions at high redshifts \citep[e.g.][]{Bouwens15b,Finkelstein15} or from cosmic infrared background radiation \citep{Mitchell-Wynne15}, whereas $\xi_\mathrm{ion}$ can be estimated from a combination of models \citep[e.g.][]{Duncan15,Stanway15,Wilkins15} and observations \citep[e.g.][]{Stark15,Bouwens15c}.

Independently assessing $\fesc$ at $z\gtrsim 6$ remains an outstanding problem, since any escaping LyC photons (rest-frame wavelength $\lambda < 912$ \AA) at these redshifts will be absorbed by the neutral IGM and hence cannot be directly detected \citep{Inoue14}. The latest studies indicate that galaxies can explain the reionization of the Universe if the galaxy population at $z\gtrsim 6$ exhibits LyC leakage at a {\it typical} level of $f_\mathrm{esc}\sim 0.1$ \citep[e.g.][]{Robertson15,Mitra15,Atek15,Stanway15}. The galaxy population at $z\approx$ 0-3, where the direct detection of LyC photons is possible, does not appear to meet this requirement \citep[e.g.][]{Bergvall13,Grazian15,Siana15,Rutkowski15,izotov2016,Leitherer16}, which means that substantial evolution in the typical LyC escape properties needs to take place at redshifts closer to the reionization epoch. 

Can we somehow determine if ionizing photons are really getting out of galaxies at $z\gtrsim 6$? Are there individual galaxies with very high levels of LyC leakage, as predicted by e.g. \citet{Kimm14,Paardekooper15} at these redshifts? If such objects exist, how do they differ from the extreme LyC leakage candidates detected at low redshifts \citep[e.g.][]{vanzella2016}? 

\citealt{zackrisson2013} argue that, since $\fesc$ regulates the relative contributions of stars and nebular emission to the spectral energy distribution (SED) at non-ionizing energies, spectroscopy with the NIRSpec spectrograph on the upcoming James Webb Space Telescope (JWST) should at least be able to constrain $\fesc$ for the brightest reionization-epoch galaxies at $z\approx$ 6--9. Assessing $\fesc$ is important also outside the context of cosmic reionization since this parameter, in shaping the SEDs of high-redshift galaxies, will also affect attempts to constrain the age, dust attenuation and stellar mass of these objects.

The SED method proposed by \citet{zackrisson2013} is complementary to the absorption line method proposed by \citet{Jones13} but has the advantage of not requiring very high signal-to-noise ($S/N$) or high spectral resolution, and should therefore be applicable to a very large number of the galaxies observed with NIRSpec during the lifetime of JWST. Moreover, a recent attempt to validate the absorption line method at $z\approx 3$ indicates unforeseen problems with this approach \citep{Vasei16}. While \citet{zackrisson2013} only discuss very simple spectral diagnostics (primarily the slope $\beta$ of the UV continuum; and the equivalent width of the H$\beta$ emission line), information on $\fesc$ will be imprinted in many spectral features throughout the rest-frame UV and optical SEDs of star-forming galaxies \citep[see Fig.~2 in][]{zackrisson2013}. Here, we will investigate the prospects of using a more extended set of spectral data for the retrieval of $\fesc$ information from JWST/NIRSpec spectra of reionization-epoch galaxies. 

The commonly used approach of deriving constraints on galaxy parameters (e.g. age, metallicity, stellar mass) from observed spectra, using a fit to model spectra weighted by the inverse of the observational
error of each data point, is unlikely to be the optimal approach in this case, since this assumes that information on the parameters of interest is intrinsically equally distributed across the spectral bins. As shown by \citet{zackrisson2013}, this does not hold for $\fesc$, which only affects selected spectral features across the rest-frame UV and optical wavelength range covered by JWST/NIRSpec for galaxies at $z\approx$ 6--9. Instead we use the lasso regression algorithm to identify the key spectral features relevant for the problem. The lasso model is trained on mock spectra of $z\approx 7$ galaxies with various levels of LyC leakage provided by the LYman Continuum ANalysis (LYCAN) simulation project \citep{Zackrisson16}, after degrading these SEDs to the spectral resolution of JWST/NIRSpec and adding observational noise relevant for realistic JWST observations.

The structure of this paper is as follows. In \autoref{sec:simulations} we describe the simulations used, and in \autoref{sec:lasso} the lasso algorithm. We then present our results in \autoref{sec:results}. We begin by discussing the results for a single simulation suite, and then go on to discuss the robustness of our model in light of various simulation uncertainties. Finally, we summarize and discuss the results in \autoref{sec:discussion}. 

\section{Simulations}
\label{sec:simulations}
In this section, we describe our simulations. The first step is to simulate the spectra of samples of galaxies during the epoch of reionization under various assumptions. The second step is to use these spectra to create mock observations for JWST/NIRSpec.

\subsection{Galaxy simulations}
In this paper, we make use of a subset of the simulated spectra of reionization-epoch galaxies described in \citet{Zackrisson16}. In short, we use the detailed star formation histories and internal metallicity distributions of  $\approx 1400$ galaxies at $z=7$ drawn from three different numerical simulations: 106 objects with stellar population masses $M_{\star}\geq 10^{7}M_{\Sun}$ \citep{finlator2013}, 874 objects with $M_{\star}\geq 10^{7}M_{\Sun}$ from CROC \citep{gnedin2014a,gnedin2014b} and 406 objects with $M_{\star}\geq 5\times 10^{8}M_{\Sun}$ from \citet{shimizu2014}.

Grids of synthetic spectra at metallicities ranging from $Z=10^{-7}$ to $Z=0.050$ generated with the Yggdrasil spectral evolutionary model \citep{zackrisson2011} were then used to compute realistic spectra with both stellar and nebular contributions for these objects. These grids were based on stellar population spectra from either Starburst99 \citep{leitherer1999} with Padova-AGB or Geneva stellar evolutionary models for non-rotating stars at metallicities $Z\geq 0.0004$, or on spectra for binary stellar populations from BPASS2 \citep{eldridge2009} at $Z\geq 0.001$. For extremely metal-poor stars ($Z=10^{-7}$--$10^{-5}$), spectra from \cite{raiter2010} were used. The associated nebular emission was computed using the photoionization code Cloudy \citep{Ferland13}, and the effect of Lyman continuum leakage was modeled under the assumption of an ionization-bounded nebula with holes free of gas and dust \citep[for details, see][]{zackrisson2013}. 

Dust was added to the Shimizu galaxies using the dust recipe described in \cite{shimizu2014} while the dust recipe by \cite{finlator2006} was used for the CROC and Finlator galaxies. These recipes -- which assign a different dust reddening $E(B-V)$ to each simulated galaxy -- were finally combined with LMC or SMC dust attenuation laws \citep{pei1992} or with the \cite{calzetti2000} attenuation law (with nebular emission being more highly affected by dust than the stellar component) to produced the final simulated galaxy spectra. In all cases, we assume that the channels through which LyC is escaping are directed away from the observer, since the model spectra would otherwise effectively return to the dust-free case at high $f_\mathrm{esc}$.

\begin{figure*}
\centering
\includegraphics[width=2.0\columnwidth]{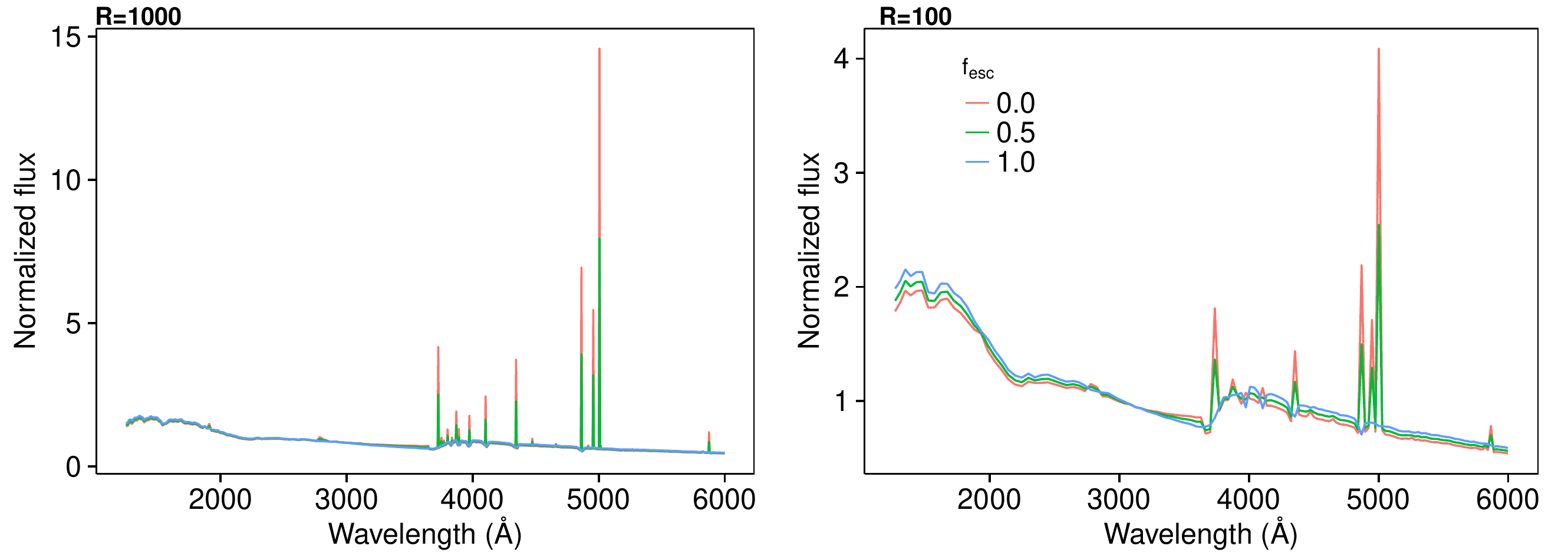}
\caption{Examples of simulated, noise-free spectra at different $\fesc$, for NIRSpec R=1000 (left) and R=100 (right) modes.}
\label{fig:sample_spectra}
\end{figure*}

We have compared the luminosity weighed average LyC photon production efficiency (according to the definition in \citealt{Wilkins15}) for our simulations with measured values by \cite{Stark15} and \cite{Bouwens15c} as well as estimates from authors including \cite{Madau99}, \cite{bouwens12a}, \cite{Finkelstein12}, \cite{Duncan15} and \cite{Wilkins15}. We find a good agreement with these values. While the \cite{Bouwens15c} \textsc{H}$\alpha$ measurements were done on galaxies between redshifts 3.8-5.4, simulations by \cite{Wilkins15} imply that the LyC production efficiency has only a weak evolution with redshift. These tests will be described in more detail in Binggeli et al (in prep).

\begin{table}
\caption{Summary of the galaxy simulations used in the paper. See the text for references.}
\label{tab:simulations}
\begin{tabular}{r|lll}
\hline \hline \\
Name & Simulation & Stellar evolution &  Dust attenuation\\
\hline \\
ShGeLMC & Shimizu & Geneva &  LMC \\
FiGeLMC & Finlator & Geneva &  LMC \\
CRGeLMC & CROC & Geneva &  LMC \\
ShPaLMC & Shimizu & Padova &  LMC \\
ShBPLMC & Shimizu & BPASS &  LMC \\
ShGeCal & Shimizu & Geneva & Calzetti \\
ShGeSMC & Shimizu & Geneva &  SMC \\
\hline
\end{tabular}
\end{table}

Using the simulations described above, we can create mock samples of galaxies using different combinations of cosmological simulations, assumptions concerning stellar evolution and dust attenuation. We will use the galaxies from the Shimizu simulation with the Starburst99-Geneva stellar population model and LMC dust (ShGeLMC) as our fiducial model when illustrating general properties of our method for inferring $\fesc$ from galaxy spectra. Later on, in \autoref{sec:model_dependency}, we will investigate the effects of different model assumptions. Here, we will use the combinations listed in \autoref{tab:simulations}.

\subsection{Simulated observations}
Our simulations give us high-resolution spectra for each of the galaxies. We wish to generate mock observations for the NIRSpec spectrograph on the JWST, and so we need to re-bin the simulated spectra to the resolution of NIRSpec, and simulate observational noise.

The NIRSpec spectrograph can be operated in three different modes with different resolutions. These resolutions are nominally R=100, R=1000 and R=2700, but in reality the resolution varies across the spectrum. Focusing on R=100 and R=1000, we use the tables on the official NIRSpec website\footnote{\url{http://www.stsci.edu/jwst/instruments/nirspec}} to calculate the exact bin widths for the different resolutions. The wavelength range is 0.6~$\mu$m--5.0~$\mu$m for R=100 and 1.0~$\mu$m--5.0~$\mu$m for R=1000. At $z=7$, these intervals correspond to 750/1250~\AA-- 6250~\AA. Here, we truncate our simulated spectra to the narrower wavelength interval of the R=1000 mode. \autoref{fig:sample_spectra} shows a few sample simulated spectra at the R=1000 and R=100 resolutions. In order to compare the shapes of the spectra rather than the intrinsic fluxes, we normalize all our spectra to have a mean flux of 1 (see \autoref{sec:results}).

\begin{figure}
\centering
\includegraphics[width=\columnwidth]{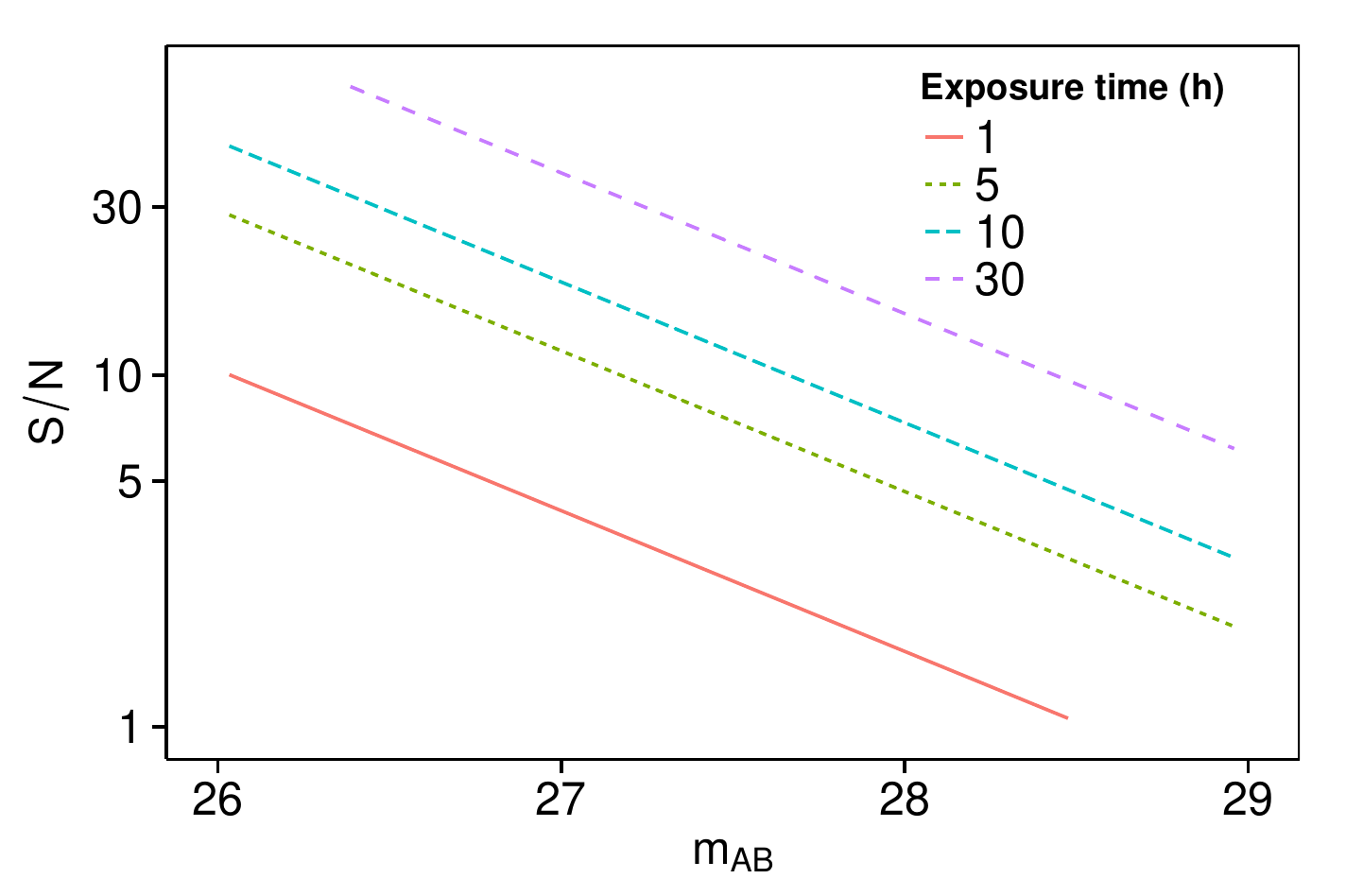}
\caption{The signal-to-noise ratio as a function of AB magnitude achievable with NIRSpec at R=100 for a bin located at 1500{\AA}. The noise levels were calculated for simulated galaxy spectra using the procedure described in the text.}
\label{fig:integration_times}
\end{figure}

The NIRSpec website also lists the minimum continuum flux observable at a signal-to-noise ratio of 10 for an exposure time of $10^4$ s, as a function of wavelength. We make the assumption that this curve scales with the square root of the exposure time in order to calculate the expected signal-to-noise at each wavelength bin for a given flux and exposure time. This assumption ignores some noise sources, such as readout noise, but should give decent approximations. This way, we can calculate the signal-to-noise ratio in each NIRSpec spectral bin for each of our simulated galaxies. We can then generate random noise realizations, making the assumption that the noise in each spectral bin is Gaussian. The sensitivity is fairly flat between 1 and 3 $\mu$m, and becomes worse for $\lambda \gtrsim 4$ $\mu$m. However, since the flux is high in the emission lines (if $\fesc$ is low), the signal-to-noise ratio is typically highest in the wavelength bins with emission lines.

Since our galaxy samples contain objects with a wide range of magnitudes, the noise for a fixed exposure time will vary greatly from object to object. In many cases, it makes more sense to show results for a fixed signal-to-noise ratio rather than a fixed exposure time. When doing so, we define the signal-to-noise of a spectrum as the signal-to-noise at the spectral bin centered at a restframe wavelength of $\lambda=1500$~{\AA}. Note that the noise will vary across the spectrum, both because of the change in detector sensitivity and because of the change in flux. \autoref{fig:integration_times} shows the signal-to-noise ratio obtained for galaxies with different AB magnitudes for different exposure times.

All the simulated spectra with and without detector noise are available for download on the LYCAN website\footnote{\url{http://www.astro.uu.se/~ez/lycan/lycan.html}}.

\section{The lasso algorithm}
\label{sec:lasso}
The aim of the  lasso (least absolute shrinkage and selection operator) algorithm \citep{tibshirani96} is to fit a linear model to a set of training data. It is similar to standard least-squares regression, but performs well even with high-dimensional data where classical approaches are prone to overfitting. 

Given $m$ training examples $(\mathbf{x_1}, y_1)$, $(\mathbf{x_2}, y_2), \ldots$ $(\mathbf{x_m}, y_m)$  with $N$ features (input variables), so that $\mathbf{x} \in \mathbb{R}^N$ and $y \in \mathbb{R}$, our aim is to find a function $\hat{y}(\mathbf{x})$ of the form:
\begin{equation}
\hat{y} = \beta_0 + \sum_{i=1}^{N} \beta_i x_{i},
\label{eq:linear_model}
\end{equation}
that best describes our data. That is, we want to obtain a set of model parameters $\hat{\boldsymbol{\beta}}$ that best predict $y$ given some value of $\mathbf{x}$.

Traditional statistical methods tend to break down and overfit the data if the number of  features, $N$, is large. The lasso algorithm solves the problem of finding $\hat{\boldsymbol{\beta}}$ in the following way:
\begin{equation}
\hat{\boldsymbol{\beta}} = \underset{\boldsymbol{\beta}}{\mathrm{argmin}} \left\lbrace \sum_{i=1}^{m} [\hat{y}(\mathbf{x}_i )- y_i]^2 + \lambda \sum_{j=1}^{N}{|\beta_j|} \right\rbrace,
\label{eq:lasso}
\end{equation}
where $\hat{y}$ is given by \autoref{eq:linear_model}. This minimization problem has no analytical solution, but can be solved numerically as a convex quadratic programming problem. The process of fitting the model on a data set is often called \emph{training}, and the data set used is called the \emph{training set}. 

\begin{figure*}
\centering
\includegraphics[width=1.9\columnwidth]{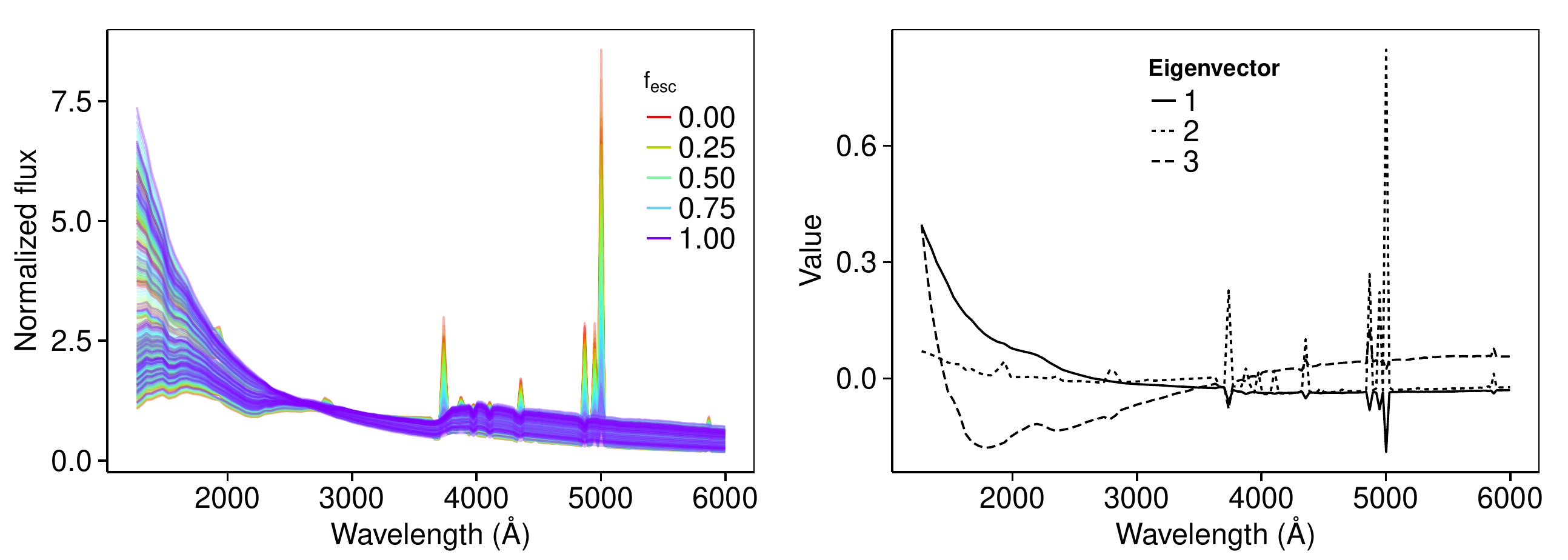}
\caption{\emph{Left panel:} spectra for 40 different sample galaxies with different assumed escape fraction from the and ShGeLMC data set. The escape fraction affects the strength of the emission lines, while the dust content gives rise to a scatter in the flux on the blue side. \emph{Right panel:} results from a Principal Components Analysis of the ShGeLMC data set. The figure shows the first three eigenvectors. Any spectrum in the data set can be approximated as a linear combination of these vectors. }
\label{fig:fesc_spectra}
\end{figure*}

In \autoref{eq:lasso}, $\lambda$ is a tunable parameter called the ``penalty'', or ``regularization parameter''. It can be thought of as a budget for the sum of the model parameters $\beta_i$. Setting $\lambda=0$ makes \autoref{eq:lasso} equivalent to standard least-squares regression. A small value of $\lambda$ gives a model that fits the test data closely, but may generalize poorly to new data points (a high variance model). Increasing $\lambda$ decreases the ``budget'', and gives a model with the less-important coefficients set to zero. A high value of $\lambda$ gives a sparser model, but a too high $\lambda$ will result in a lower predictive power (high bias). 

The fact that some coefficients become exactly zero is a distinctive property of the lasso, and a consequence of the use of the L1-norm in the second term in \autoref{eq:lasso}. Other related algorithms, such as ridge regression, do not have this property. In general, the lasso is  well suited for situations in which we expect a large fraction of the input features to be irrelevant for predicting the output variable. In our application, the input features are the fluxes in each spectral bin. Since we expect most spectral bins to have little or no correlation with $\fesc$, the lasso would seem like a suitable choice of algorithm.

When applying the lasso to a problem, an important task is to determine the best value of $\lambda$. By ``best'', we typically mean the value that gives the smallest error (defined by some error measure such as the mean of the squared errors, MSE) on new data, i.e.\ data outside the training set. This value will be different depending on the problem at hand. In some situations, a large fraction of the features will be relevant in predicting $y$, and $\lambda$ should be kept small. In other cases, we may have several features that add little but noise. In this latter situation,  a high $\lambda$ will ensure that the coefficients of these features are zero.

A common method for estimating the best value of $\lambda$ is \emph{cross-validation}, where the full data set is split into two: a training set and a (smaller) validation set. A series of models are then fit to the training set using \autoref{eq:lasso} with different values of $\lambda$ and the MSE is calculated on the validation set for each model. Typically, one finds that the cross-validation MSE is high at very large values of $\lambda$, where the model under-fits the data. It will then decrease as $\lambda$ decreases, and finally increase again at low values of $\lambda$, due to over-fitting. Normally, the $\lambda$ that gives the smallest cross-validation MSE will be the one to use, but sometimes one may opt for a slightly larger value in order to obtain a model that is easier to interpret (with fewer non-zero coefficients).

A downside of the cross-validation approach described above is that part of the data must be excluded from the training set. Therefore, it is common to perform so-called $k$-fold cross-validation. Here, the training set is is split into $k$ subsets (a common value for $k$ is 10). Then, $k$ different models are fit using the same $\lambda$, each one using $k-1$ of the subsets for training and the remaining subset for cross-validation. The cross validation error is then taken to be the mean of the errors for the $k$ different subsets.

\section{Results}
\label{sec:results}
In this section, we first investigate which parts of the spectra contain the most information about the escape fraction. We then go on to demonstrate the results of fitting a lasso model to the simulated spectra. Finally, we look at the effects of detector noise and galactic dust on our results.

\subsection{The effects of the escape fraction on galaxy spectra}
\label{sec:fesc_effects}
Before we start applying the methods described above to our simulated spectra, we take a closer look at how galaxy spectra are affected by changes in the escape fraction. The top-left panel of \autoref{fig:fesc_spectra} shows the spectra for 40 different galaxies from the ShGeLMC simulation. For each galaxy we show the spectrum for a range of $\fesc$. The galaxy spectra differ in two major ways. First, there is a spread in flux on the blue side (low wavelength) due to the varying amount of dust attenuation. Second, the strengths of the nebular emission lines differ due to the differences in the escape fraction.

To better illustrate where the information is located in the spectra, we show the results of a Principal Components Analysis (PCA) in the right panel of \autoref{fig:fesc_spectra}. The PCA procedure finds a set of orthogonal basis vectors such that the first vector is the direction of highest variance, i.e. the vector that will give the highest variance among the data points after projection to this vector. The subsequent vectors maximize the variance after subtracting the projections of the data along the previous vectors. The first few vectors found from PCA will thus explain most of the variance in the data, and can be thought of as a measure of which features in the data set carry the most information. 

From the PCA vectors, we see that most of the variance is indeed contained in the wavelength bins with strong emission lines. However, it also becomes clear that some information can be obtained from the blue side of the spectra. The eigenvalues corresponding to the eigenvectors in \autoref{fig:fesc_spectra} give an indication of how much ``information'' can be explained by the eigenvectors. In the absence of detector noise, the three eigenvectors shown in \autoref{fig:fesc_spectra} are enough to explain $\gtrsim 99$ \% of the variance in the data.

\subsection{Regression results}
\label{sec:regression_results}
In this section we present the results of using lasso regression to predict the values of the escape fraction. We begin with the simplest case, where the galaxy model is assumed to be known. For illustration purposes, we will adopt ShGeLMC as the fiducial simulation (see \autoref{sec:simulations} for details). We use the \texttt{glmnet}\footnote{\texttt{https://cran.r-project.org/web/packages/glmnet/}} package with 10-fold cross-validation to perform the minimization in \autoref{eq:lasso} and fit the model in \autoref{eq:linear_model}. 

\begin{figure*}
\centering
\includegraphics[width=1.9\columnwidth]{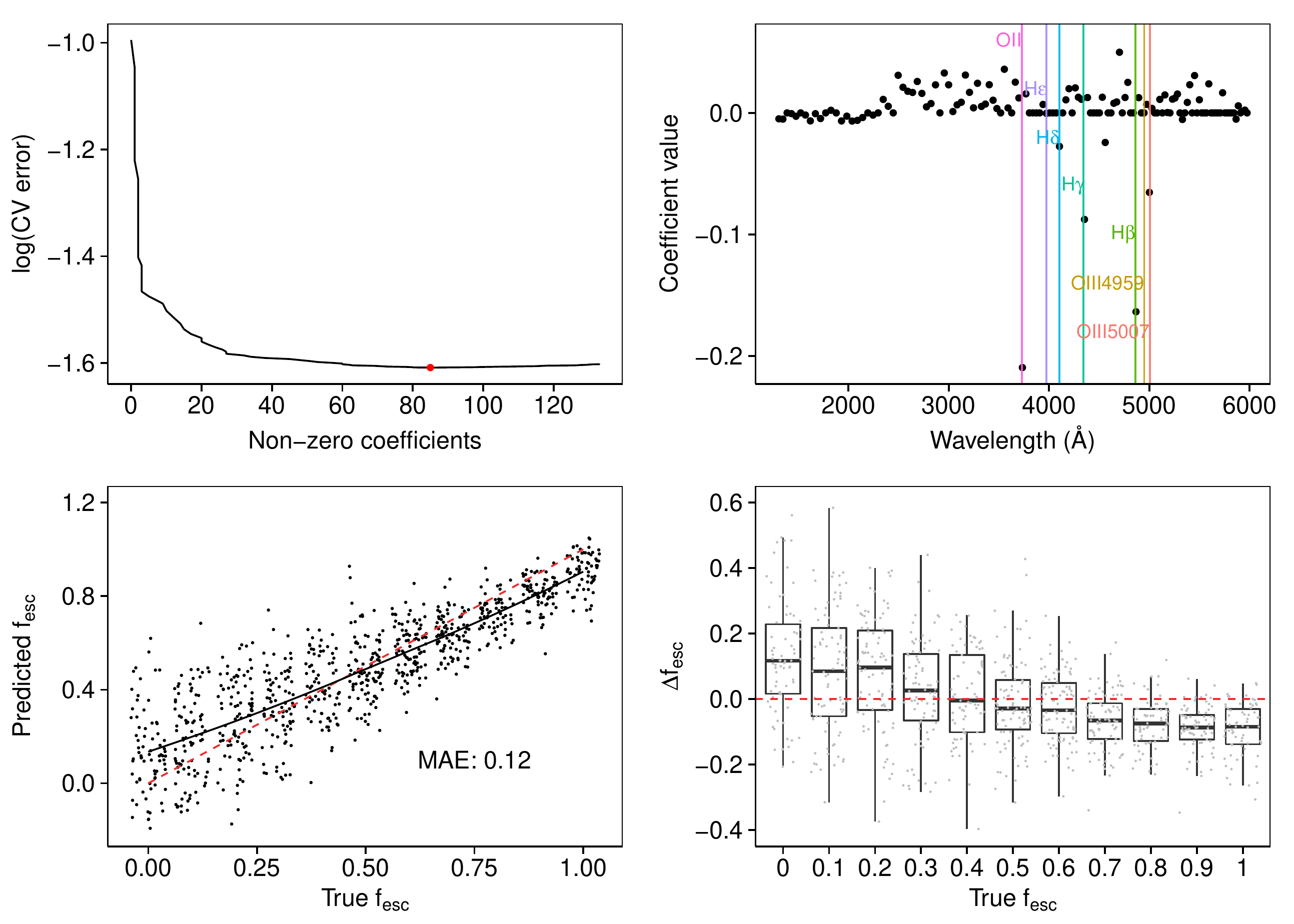}
\caption{Results when fitting a lasso regression model to the ShGeLMC simulation set at $S/N=5$. \emph{Top left}: The cross-validation (CV) error as a function of the number of non-zero coefficients in the model. The red dot indicates the number that gives the smallest cross-validation error. \emph{Top right}: The values of the model coefficients ($\beta_i$ in \autoref{eq:linear_model}) for the model with the smallest cross-validation error. The locations of some prominent nebular emission lines are shown as colored lines. \emph{Bottom left}: Predicted vs true escape fractions for the data in the test set (i.e.\ data not used to when training the model). The dotted red line shows $\fesc^{\mathrm{pred}}=\fesc^{\mathrm{true}}$, while the solid black line shows the mean of the predicted escape fractions. For visual clarity, we have added some random jitter to the values of $\fesc^{\mathrm{true}}$; in reality these are all located exactly at 0, 0.1, 0.2 etc. \emph{Bottom right}: Box-and-whiskers plot of the residuals at each $\fesc^{\mathrm{true}}$. The boxes are located at the first and third quartiles, and the whiskers extend to the highest and lowest values within 1.5 times the inter-quartile range.}
\label{fig:shimizu_geneva_peilmc}
\end{figure*}

The dataset for the ShGeLMC simulation consists of 406 simulated galaxies, each with spectra calculated for $\fesc=0, 0.1, 0.2 \ldots 1.0$  for a total of 4466 objects. The features in the model (i.e.\ $\mathbf{x}$ in \autoref{eq:linear_model}) are simply the fluxes at each NIRSpec wavelength bin, after normalizing each galaxy spectrum to have mean 1 (since we normalize the spectra, the luminosities of the galaxies have no effects on the results). The coefficients $\beta_i$ in \autoref{eq:linear_model} given by \autoref{eq:lasso} are thus the weights given to each wavelength bin, and the estimate of the escape fraction is given as the sum of the fluxes multiplied by their respective coefficients.

The results when fitting the lasso model are shown in \autoref{fig:shimizu_geneva_peilmc}. The top-left panel shows the cross-validation error as a function of the regularization parameter $\lambda$, or, equivalently, the number of non-zero coefficients in the model. We see that the best model is given by using approximately 85 of the wavelength bins. If the regularization parameter is decreased further, more coefficients are included in the model, and we start to overfit the data. The optimal value of $\lambda$ will depend on several factors, including the noise level. A higher noise level will be more sensitive to overfitting, and the best model will include a lower number of non-zero coefficients.

The top-right panel shows the values of the coefficients, i.e.\ $\beta_i$ for the regularization parameter value that gives the lowest cross-validation error. The coefficients with the highest absolute values are all located at wavelength bins containing prominent emission lines (a few of the strongest lines are marked with vertical lines). These coefficients are all negative, indicating that strong emission lines are a sign of a low $\fesc$. 

In the bottom row we show the results when applying the method to a test set, which was constructed by randomly selecting 20 \% of the objects in the simulation and removing them from the training set. Thus, the algorithm had never ``seen'' these objects before. The bottom left panel shows the predicted vs true $\fesc$ on the test set. The bottom-right panel shows a box-and-whiskers plot of the residuals, $\Delta \fesc$, for each value of the true $\fesc$. In general, the model performs better for high escape fractions than for low ones. In particular, it appears that objects with $\fesc \lesssim 0.2$ are somewhat problematic.

The mean absolute error for the predicted $\fesc$ is 0.12. As a comparison, \cite{zackrisson2013} proposed that the escape fraction of a galaxy could be inferred using two parameters derived from its spectrum: the UV slope $\beta$ and the equivalent width of the H$\beta$ line. Fitting a regression model using only these two parameters to the same training set as used for \autoref{fig:shimizu_geneva_peilmc} gives a mean absolute error of 0.16 on the same test set. For simulations with a lower dust content, the increase in performance from using the entire spectrum is even greater.

From the bottom panels in \autoref{fig:shimizu_geneva_peilmc} it is clear that there is a considerable scatter in $\fesc^{\mathrm{pred}}$, especially for low values of $\fesc^{\mathrm{true}}$. This implies that a very accurate determination of $\fesc$ will be impossible for a single object. However, the mean $\fesc$ of a sample of galaxies may still be possible to measure to a much greater accuracy, as we discuss in more detail in \autoref{sec:mean_fesc}.


It is interesting to investigate in some more detail how the model can be interpreted. From the top-left panel of \autoref{fig:shimizu_geneva_peilmc} we see that the cross-validation error drops very sharply when the first few coefficients are added, indicating that most of the information about the escape fraction is contained in a small number of wavelength bins. Adding more coefficients to the model only results in a modest increase in performance. Which wavelength bins are the most important? From \autoref{fig:fesc_spectra}, we would suspect that the first coefficients to be added are associated with the most prominent emission lines. Varying the regularization parameter reveals that this is indeed the case---a high value of $\lambda$ in \autoref{eq:lasso} results in a model containing only the emission lines. 

If $\lambda$ is decreased, coefficients on the blue end of the spectrum are added, followed by wavelength bins in-between the major emission lines. Some of these latter bins actually contain very weak lines---sometimes more than one line per bin---that are buried in noise and invisible in the noisy spectra. However, seen over the average of the entire data set, they do contain small amounts of information. The interstellar dust changes the slope of each spectrum. Since we normalize all our spectra to have the same mean value, a spectrum with a shallow slope will get a higher relative flux in the red part of the spectrum, where the emission lines are. The lasso model automatically corrects for this by adding positive coefficients in-between the emission lines. These coefficients ensure that the emission line strengths are measured relative to the continuum level, thus correcting for the effects of the dust.

In \autoref{sec:bootstrap}, we show the results from a bootstrap analysis of the robustness of the model coefficients.

\subsubsection{Noise level and spectral resolution}
In the results shown above, the model was trained on spectra with a noise level of $S/N=5$. In a real-world survey, the signal-to-noise will depend on the brightness of each individual galaxy, and thus each object will have a different signal-to-noise ratio. We may ask which noise level should be used when training the model. To answer this question, we created data sets with simulated NIRSpec noise at various levels and fit lasso models to each noise level. We then evaluated the performance of each of these models on a series of test sets with different noise levels.

\begin{figure}[htb]
\centering
\includegraphics[width=\columnwidth]{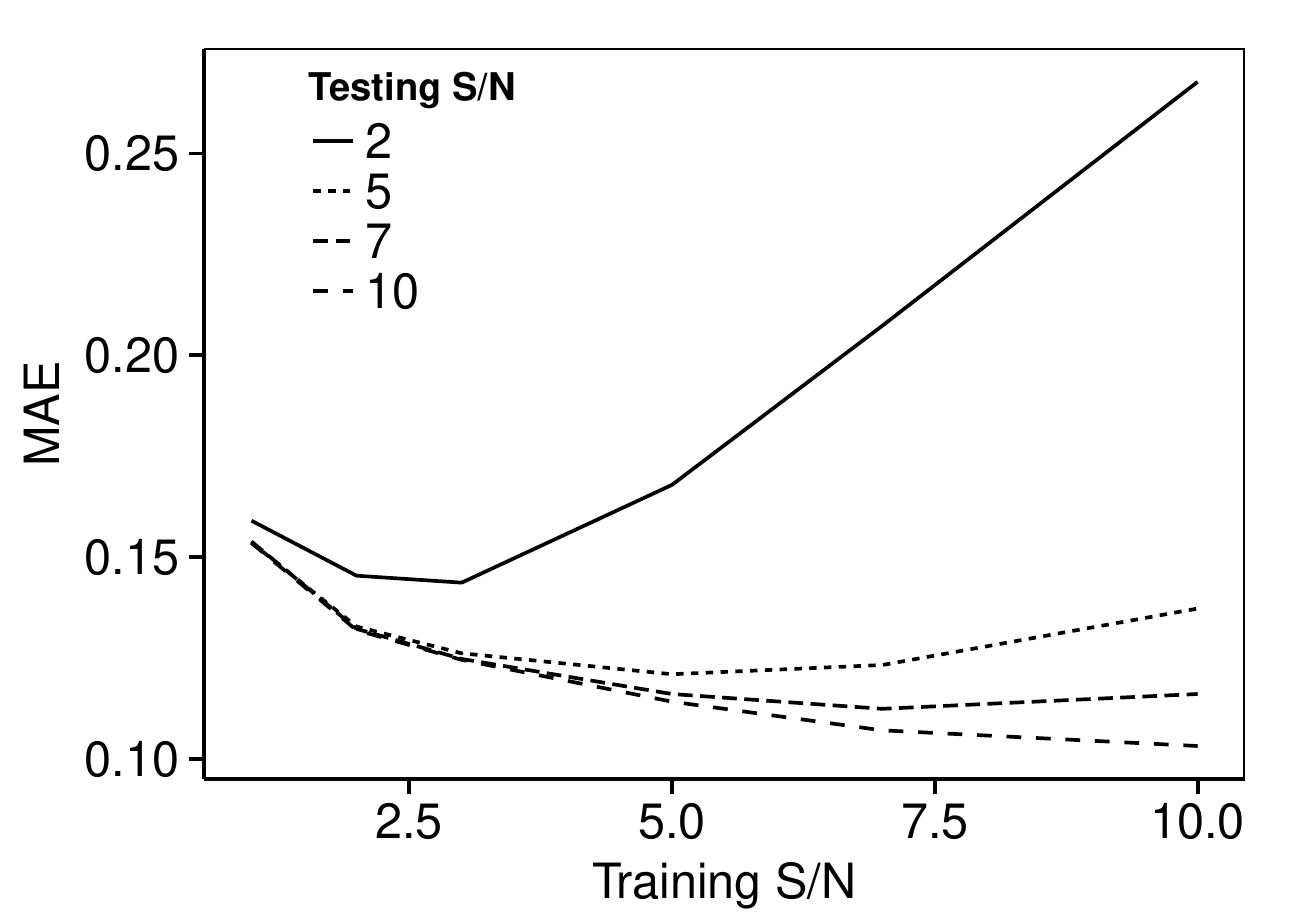}
\caption{The mean absolute error when varying both the test set and training set noise level. In general, the best performance is obtained when the training set has a noise level similar to that of the test set. All results in this figure are for the ShGeLMC simulation.}
\label{fig:vary_sn}
\end{figure}

The results of this test are shown in \autoref{fig:vary_sn}. Each of the lines in this figure show the results when holding the signal-to-noise ratio in the test set constant while varying the noise level in the training set. We see that the lines all have minima approximately where the training set noise level equals the test set noise level, suggesting that the best performance is obtained when the model is trained on data with a similar noise level to the data it will be applied to. The reason for this is that  a model that is trained on noise-free or low-noise data will tend to rely on features such as weak emission lines that may be strongly correlated with $\fesc$, but will be easily obscured by noise. Training on high-noise data, on the other hand, tends to produce a simpler model that only accounts for the strongest emission lines. If we apply a model trained on one noise level to data with a different noise level, the model will either omit information that is in fact measurable, or try to make use of information that is buried in noise.

A related question is what spectral resolution is most effective for predicting the escape fraction. The NIRSpec instrument can be operated in several different resolution modes. With a higher resolution, one can discern finer details in the spectra, but the noise in each wavelength bin will also be higher.

\begin{figure}[htb]
\centering
\includegraphics[width=\columnwidth]{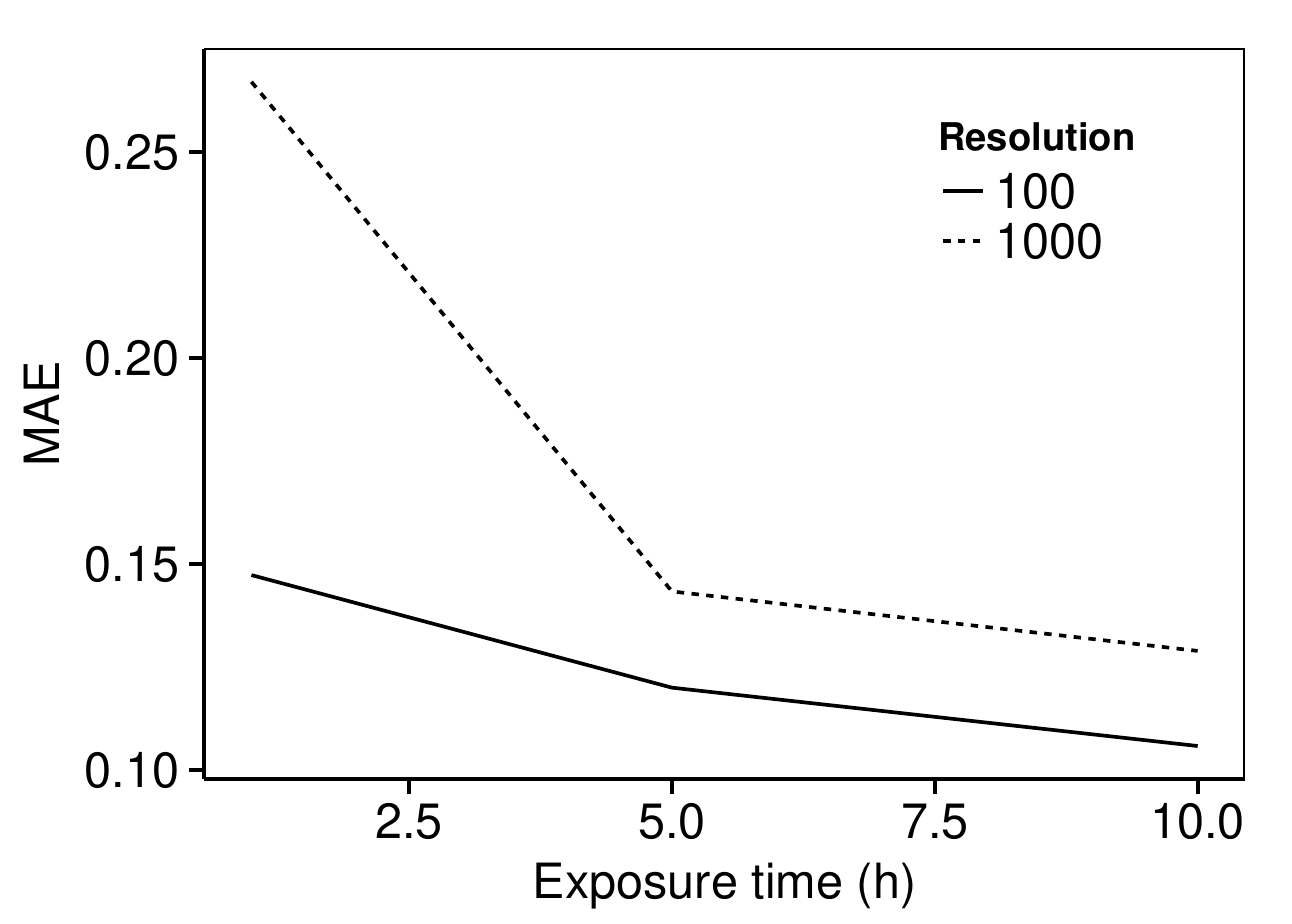}
\caption{The mean absolute error as a function of exposure time for galaxy spectra observed with NIRSpec resolution modes R=100 and R=1000 respectively. All results are for the ShGeLMC simulation. The actual values on the $y$ axis depend on the brightnesses of the galaxies in the simulated sample, which do not correspond to what would be observed in a realistic survey. Only the difference between the two curves is meaningful.}
\label{fig:resolution_comparison}
\end{figure}

In \autoref{fig:resolution_comparison} we show the mean absolute error on the test set when training a model on the ShGeLMC simulation at NIRSpec resolutions R=100 and R=1000 as a function of exposure time. Fixing the exposure time will result in a different signal-to-noise ratio for each object, since the data set contains galaxies with a range of different brightnesses. Since the objects come from a large simulation volume, this range of brightnesses does not correspond to what would be observed in a real NIRSpec survey, but we can nevertheless compare the results for the two different resolutions.

The R=100 results are consistently better than the R=1000 results for a given exposure time. While the R=1000 spectra will contain more information, it would seem that the additional noise in the high-resolution spectra degrades the results to such a degree that the additional information is not useful. We conclude that R=100 is the better resolution to use for our method. For the remainder of the paper, we will show only results for R=100.

\subsection{Dependence on simulation parameters}
\label{sec:model_dependency}
In the previous section, we showed that lasso regression can be used to construct a model that will predict $\fesc$ from a NIRSpec spectrum with decent precision, especially for higher values of $\fesc$. However, all our results are based completely on simulations, and thus they are dependent on the various assumptions that go into the simulations. This will remain true even after real observations are obtained, since there is no way to directly measure the escape fraction of galaxies during the epoch of reionization.

In this section, we explore the sensitivity of our results to simulations and model assumptions. As discussed in \autoref{sec:simulations}, we can roughly divide the process for generating our mock spectra into three parts: galaxy simulation, stellar evolution model and dust attenuation model. We consider galaxies extracted from simulations by Shimizu, CROC and Finlator, using Geneva, Padova and BPASS2 stellar evolution models. For the dust modeling, we take the dust content from the galaxy simulations and assume Calzetti, Large Magellanic Cloud (LMC) or Small Magellanic Cloud (SMC) attenuation models. In total, this leaves us with $3\times 3 \times 3 = 27$ possible combinations.

\begin{figure*}[htb]
\centering
\includegraphics[width=2\columnwidth]{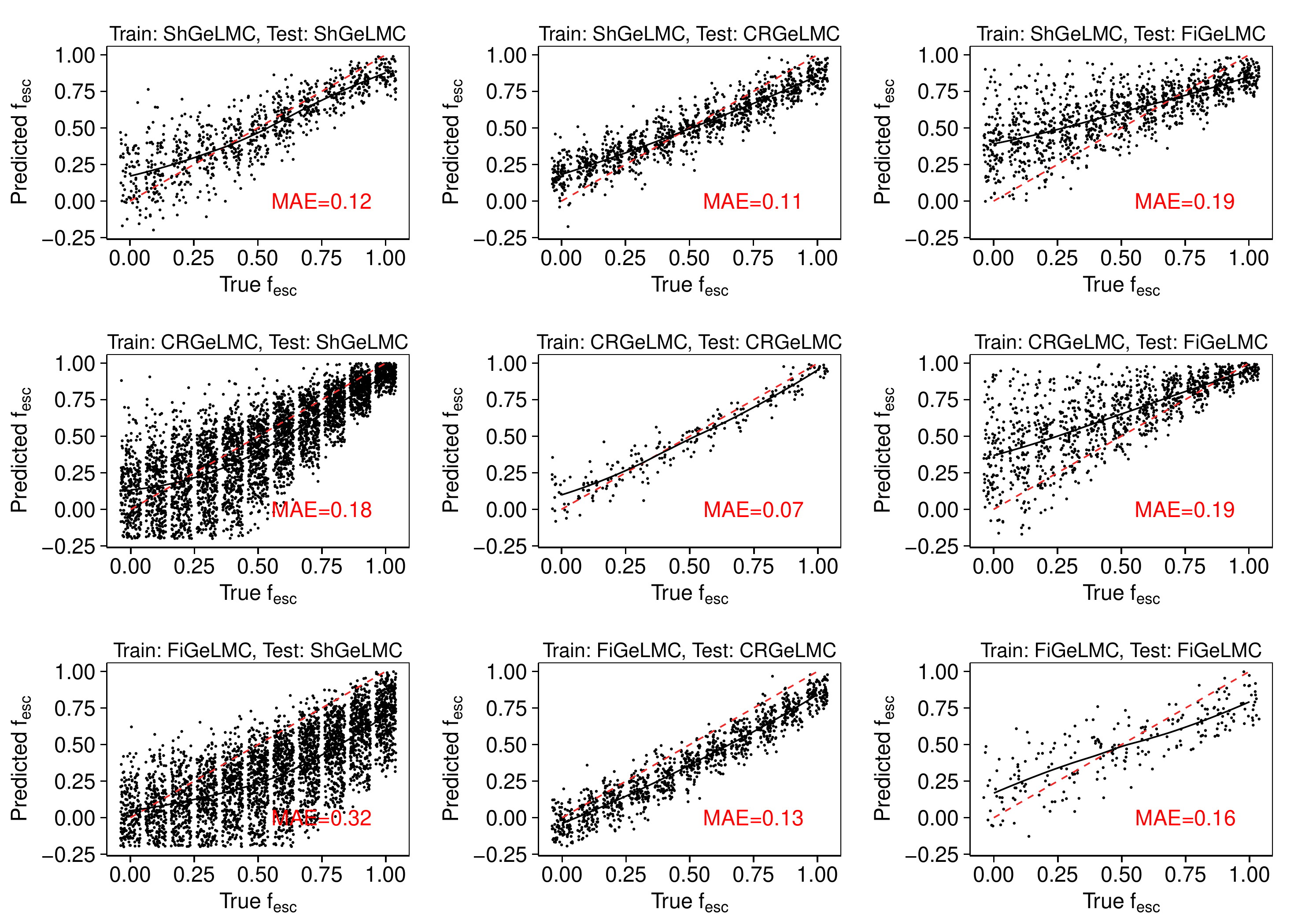}
\caption{Results when applying a regression model trained on a particular simulation to a different simulation. Here, we show the results for the Shimizu, CROC and Finlator simulations, all using Geneva stellar evolution models, and LMC dust attenuation. For the panels along the diagonal, where the test simulation is the same as the training simulation, the test set was constructed by randomly drawing 20~\% of the objects from the training set (these objects were not used for training).}
\label{fig:vary_simulation}
\end{figure*}

We can now ask ourselves what will happen if we train our model on data produced using one set of assumptions, and use it to predict the escape fractions of galaxies produced using a different set of assumptions. Rather than showing the results of all $27 \times 26 =702$  training/testing model combinations, we limit ourselves to varying one component (galaxy simulation, stellar evolution model, dust attenuation model) at a time.

\autoref{fig:vary_simulation} shows the results of varying the galaxy simulation, while using the Geneva stellar evolution model and LMC dust attenuation. For example, the middle panel in the top row shows the results of training a lasso regression model on galaxies from the Shimizu simulation and applying it to galaxies from the CROC simulation. Perhaps what's most striking in this figure is the large errors when applying models trained on the Shimizu or CROC simulations on galaxies from the Finlator simulations (right column). The reason for the large scatter at low escape fractions is the much greater range of star formation histories in the Finlator galaxies, which causes a greater spread in the shapes of the spectra. This is problematic when attempting to apply a model trained on a more homogeneous data set, such as Shimizu or CROC. Furthermore, the Finlator data set contains only one-fourth the number of objects of the Shimizu data set, which further reduces the performance of the algorithm.

\begin{figure*}[htb]
\centering
\includegraphics[width=2\columnwidth]{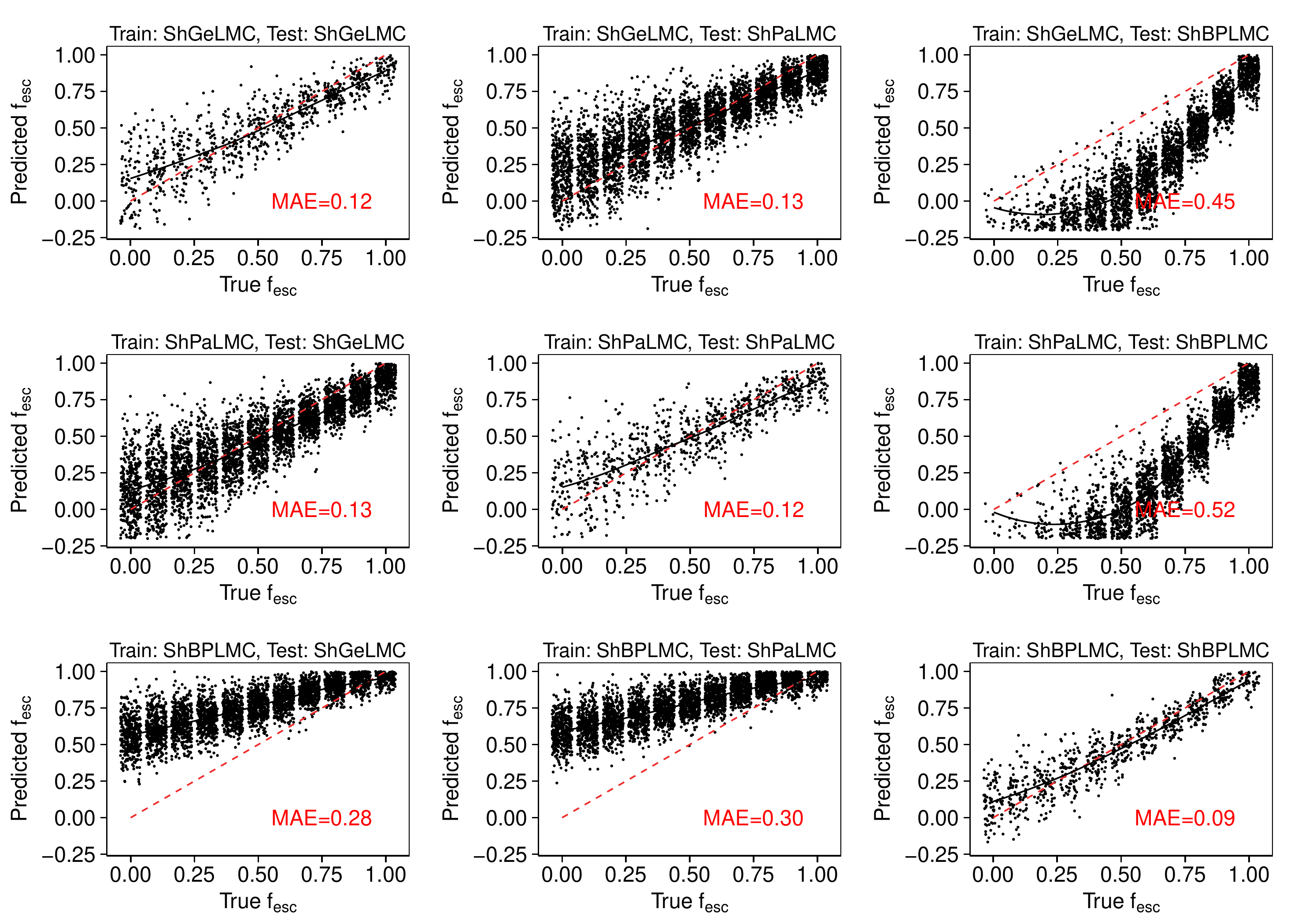}
\caption{Same as \autoref{fig:vary_simulation}, but varying the stellar evolution model. All data sets are from the Shimizu simulation, assuming LMC dust attenuation.}
\label{fig:vary_stellar_track}
\end{figure*}

In \autoref{fig:vary_stellar_track}, we show the results when assuming the wrong stellar evolution model. Here, we use only galaxies from the Shimizu simulations, assuming LMC dust, calculating the results for the Geneva, Padova and BPASS2 stellar evolution models. The worst issues occur with models using BPASS2. Models that are trained on BPASS2 overestimate $\fesc$ when applied to Geneva or Padova models, whereas models trained on Geneva or Padova underestimate $\fesc$ when applied to BPASS2 data. The reason for this bias is that BPASS2 predicts stronger emission lines for a given escape fraction than Geneva or Padova. Therefore, a model trained on BPASS2 will expect low-$\fesc$ galaxies to have stronger emission lines than what is possible according to Geneva or Padova. 

\begin{figure*}[htb]
\centering
\includegraphics[width=2\columnwidth]{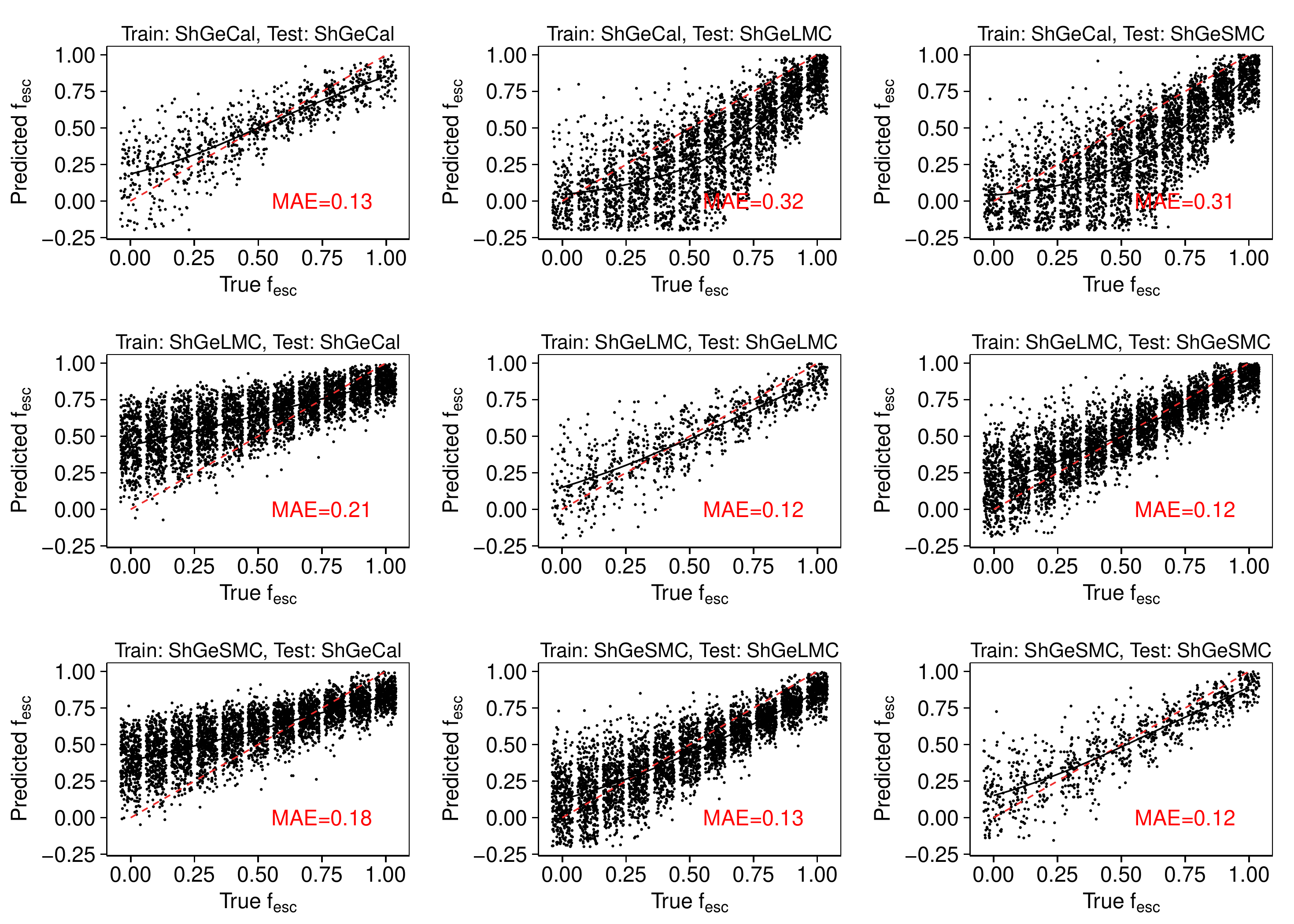}
\caption{Same as \autoref{fig:vary_simulation}, but varying the dust attenuation. All data sets are from the Shimizu simulation, using the Geneva stellar evolution model.}
\label{fig:vary_dust}
\end{figure*}

Finally, we show the results of varying the dust model in \autoref{fig:vary_dust}. We see that the difference between the LMC and SMC attenuation models produce negligible differences in the models, which is to be expected, since these two models are very similar. However, models trained on data using the Calzetti attenuation law---which differs significantly from SMC and LMC---are heavily biased when applied to the other models.

\subsection{How useful are the results?}
\label{sec:mean_fesc}
As we have seen above, even when the galaxy simulation is assumed to be accurate, our model gives large errors when $\fesc^{\mathrm{true}}$ is low. This is potentially problematic, since the typical $\fesc$ is expected to be only around 0.1--0.2 at $z=7$, as discussed in the introduction. Judging by, for instance, \autoref{fig:shimizu_geneva_peilmc}, it may appear that $\fesc=0.2$ is indistinguishable from $\fesc=0$. However, that does not necessarily mean that no useful information can be obtained using our method.

First, while the \emph{typical} escape fraction may only be around 0.1--0.2, high-$z$ galaxies will likely have a distribution of escape fractions, with some objects having significantly higher values. Our method could be used to identify these outliers, which would be of great interest to the study of reionization.

Second, while it may be impossible to reliably separate an individual $\fesc= 0.2$ galaxy from an $\fesc=0$ galaxy, the situation will be quite different if we consider instead the mean escape fraction for a population of galaxies. To test how accurately we can determine the population mean of $\fesc$, we applied the same model as in \autoref{fig:shimizu_geneva_peilmc} to one test set with only objects with $\fesc^{\mathrm{true}}=0$ and one with $\fesc^{\mathrm{true}} = 0.2$. From these test sets, we used the bootstrap method \citep{efron1979} to draw 500 subsamples of a fixed size $n_{\mathrm{gal}}$. We then used our model to predict $\fesc$ for each object, and calculated the mean $\fesc^{\mathrm{pred}}$ for each subsample.

\begin{figure}[htb]
\centering
\includegraphics[width=\columnwidth]{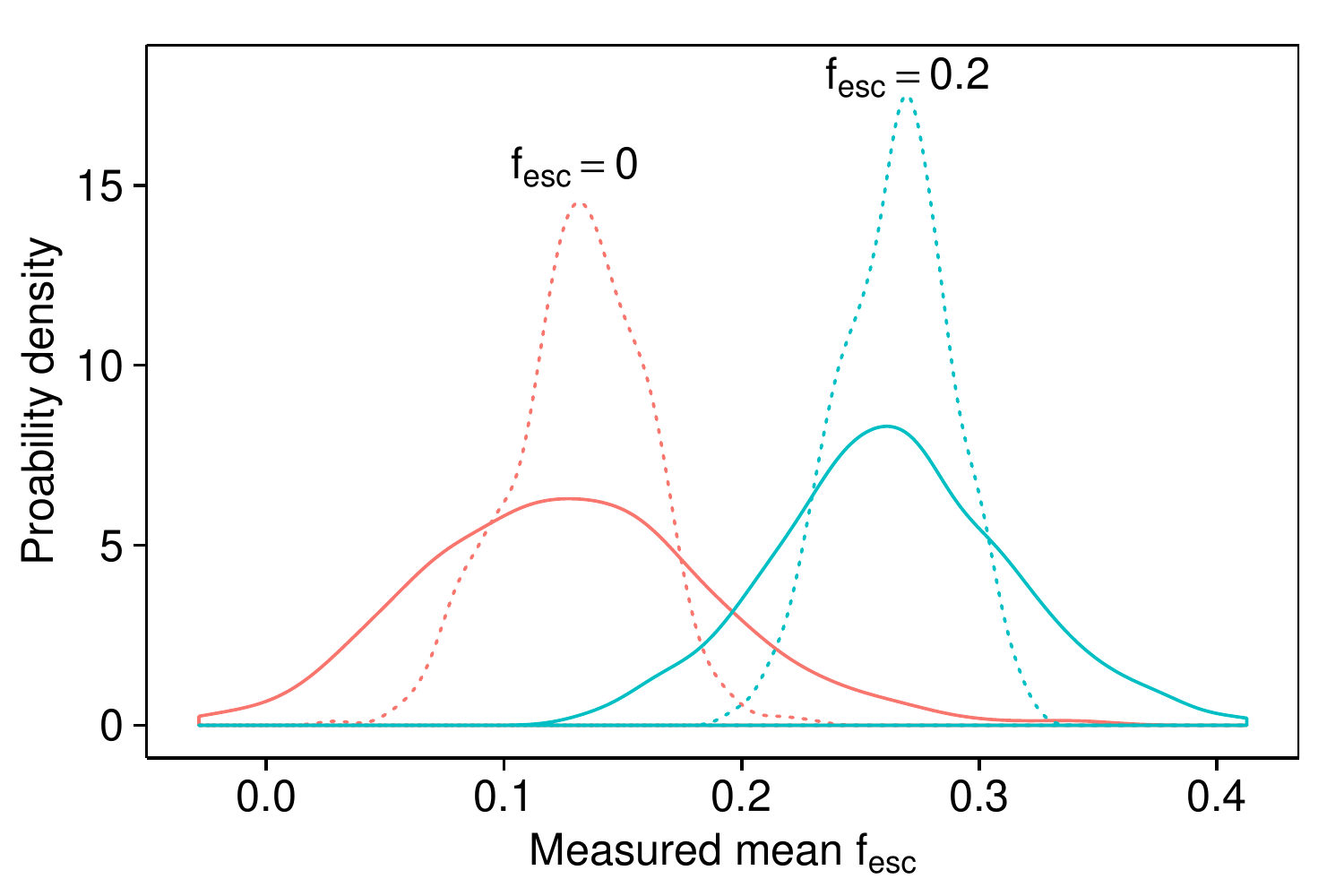}
\caption{Results from measurements of the mean escape fraction for a sample of 10 (solid lines) and 50 (dotted lines) galaxies. The curves show the distributions of the mean measured escape fraction for 500 bootstrap samples, given an underlying population with $\fesc^{\mathrm{true}}=0$ (red lines) and $\fesc^{\mathrm{true}}=0.2$ (cyan lines).}
\label{fig:bootstrap_mean_fesc}
\end{figure}

\autoref{fig:bootstrap_mean_fesc} shows the distribution of mean predicted $\fesc$, for samples of size $n_{\mathrm{gal}}=10$ (solid lines), and $n_{\mathrm{gal}}=50$ (dotted lines). The red lines are for the case where $\fesc^{\mathrm{true}}=0$, and the cyan lines are for $\fesc^{\mathrm{true}}=0.2$. The curves in \autoref{fig:bootstrap_mean_fesc} thus essentially indicate the probability of measuring a given mean $\fesc$ in the case where the entire population has $\fesc=0$ or $\fesc=0.2$. While the bias in the model, discussed above, is evident, the results still look promising when it comes to distinguishing between populations with different escape fractions. The $n_{\mathrm{gal}}=50$ curves (dotted) are very well separated, indicating that for a sample size of 50, we can tell $\fesc=0.2$ apart from $\fesc=0$ with high confidence. For $n_{\mathrm{gal}}=10$, there is more overlap between the curves, but it may still be possible to tell the two cases apart. For a situation where $\fesc^{\mathrm{true}}=0.1$ everywhere (not shown in the figure), we require a sample size of $n_{\mathrm{gal}}\approx100$ to reliably measure a mean escape fraction that is distinct from zero.

The discussion above is simplified in a number of ways. First, we assume that the individual members of the galaxy population all have the same escape fraction, i.e.\ the distribution of $\fesc$ is a delta function. As we noted previously, this is not likely to be true. Second, we assumed that all galaxies in our mock samples could be observed at the same signal-to-noise. A realistic sample would contain galaxies with a range of luminosities, and would therefore have a different noise level for each object. Nevertheless, this simple test shows that measuring the mean escape fraction of a large sample is possible even if the errors are large for individual objects.

\section{Summary and discussion}
\label{sec:discussion}
In this paper, we have shown how machine learning algorithms such as lasso regression can be applied to simulated galaxies to obtain a function that maps a spectrum to a Lyman continuum escape fraction. Our method uses the entire observable spectrum, which leads to an increase in performance over, for instance, using only the UV slope and the H$\beta$ line \citep{zackrisson2013}. 

We find that using the lowest NIRSpec resolution, R=100, gives the best results for a given exposure time. Interstellar dust presents a challenge for our method, but it can be dealt with unless the dust attenuation law is very poorly known. Training a model on data with low noise results in a more complicated model that uses a larger number of spectral bins compared to a model trained on data with high noise. The performance of the model is always better when the training and testing data have the same noise level. Therefore, in a real-world situation, a suitable approach might be to train several models using different noise levels, and apply them depending on the noise level for each specific observed galaxy.

In a sense, this method represents the reverse of the more traditional approach used when fitting spectral energy distributions (SEDs), which has also been used to constrain $\fesc$ \citep{ono2010}. In SED-fitting, one usually generates a large grid of model galaxies, and looks for a model galaxy that looks similar to a given observation. It is then assumed that the properties of the observed galaxy will be similar to that of its closest match in the model grid. 

The problem with this approach lies in defining ``similar''.  Typically, the similarity of two SEDs is measured simply as mean of the squared difference between the fluxes in all filters. However, when constraining a single property, such as $\fesc$, not all wavelength regions are equally important. In fact, our results show that most of the information about $\fesc$ comes from only a few wavelength bins. The method presented here automatically finds the optimal way of comparing observations to simulations, when the goal is specifically to infer the escape fraction.

We emphasize, however, that this is not a model-independent method. Like all other indirect approaches to determining $\fesc$, our method can only be as reliable as the simulations it is based on. In section \ref{sec:model_dependency} we demonstrated this by applying the algorithm to a set of simulations different from those which it was trained on (a situation similar to analyzing real galaxies that have properties that are systematically different from those in the training set). 

These tests reveal a few different complications. For instance, assumptions about stellar binarity can significantly alter the prediction. Since the BPASS2 model (which assumes binary stellar evolution) produces significantly stronger nebular emission lines than the Geneva or Padova models (which are based on single-star evolution), this leads to a bias on the inferred $\fesc$ in cases where the algorithm has been trained on models with binary star properties very different from those prevalent at $z>6$. It is likely that similar biases can be produced by incorrect assumptions concerning stellar rotation \citep[e.g.][]{Topping15} or the inclusion of stars significantly more massive than $100 M_\odot$ \citep[e.g.][]{Stanway15}. 
Methods aiming to test or calibrate spectral synthesis models at both high and low redshift \citep[e.g.][]{Bouwens15c,Wofford16,Steidel16} will therefore be crucial for the success of our proposed method.

Assumptions about the dust attenuation law has similar effects but for somewhat different reasons. The attenuation law affects the UV slope, and our model automatically compensates for this when measuring the relative strengths of the emission lines. However, this compensation will not work if the assumed attenuation law is a poor representation of that actually prevalent in $z=7$ galaxies.

Furthermore, the simulations used in this paper all predict a fairly narrow distribution of star formation histories in the galaxies sufficiently massive to be within range of JWST/NIRSpec. If actual galaxies in the reionization epoch display greater variety, this could also bias the inferred $\fesc$. Extreme quenching, in which a period of intense star formation is followed by no star formation at all, is not seen in the simulation suites used here but would be particularly troublesome for our method. In the post-starburst phase, such galaxies could briefly exhibit blue $\beta$ slopes while having very low LyC production and therefore no emission lines, thereby mimicking star-forming galaxies with high $\fesc$.

The method should ideally be trained on mock spectra based on both LyC leakage mechanisms discussed by \citet{zackrisson2013} -- ionization-bounded nebulae with holes and matter/density-bounded nebulae -- but since the latter case involves more free parameters, this is saved for future work.

For the entire paper, we have presented our results in terms of fixed signal-to-noise ratios. How many galaxies can we expect to observe with a given signal-to-noise ratio in a future JWST/NIRSpec survey? From \autoref{fig:integration_times}, we see that a magnitude brighter than  $m_{\mathrm{AB}}\approx 29$ is needed to get $S/N \gtrsim 5$ within approximately 10 hours of exposure time. Taking the UV luminosity function at $z=7$ from \cite{Bouwens15b} we find that there should be around 100 such objects (which is close to the number of microshutters available) in a single NIRSpec pointing ($3.4 \times 3.4$ arcmin), for $\Delta z=1$. 

This number depends on the faint end of the luminosity function which is still somewhat uncertain at these redshifts. However, it seems reasonable to expect that it will be possible to obtain decently-sized spectroscopic samples at $z\approx 7$, for which our method can be applied, especially if multiple NIRSpec pointings are used. Furthermore, even a sample with a lower signal-to-noise ratio may be useful if the number of objects is large enough. Even if $\fesc$ can only be determined with a significant uncertainty for a single object, it may still be possible to infer the mean value of the population, which is the most important quantity for reionization studies. A possible caveat here is that $\fesc$ may be highly dependent on galaxy mass (or luminosity), with the majority of ionizing photons coming from faint sources, below the detection limit of even JWST (e.g.\ \citealt{Atek15}.

The most important factor for improving the reliability of our method is to obtain more reliable simulations. There are still significant uncertainties regarding the properties of high-redshift galaxies, and we have seen that different assumptions regarding, for example, dust attenuation and stellar evolution can lead to different predictions for the escape fraction. Hopefully, as new observational data becomes available over the next few years, simulations can be better calibrated to the physical processes at high redshifts.

There is also some room for improvement in the method itself. We have seen that there is some bias in our predictions, especially for the Finlator simulations which contain a wider range of galaxy properties. It is possible that this bias may be reduced by extending the linear model used here with a suitable link function, for example a logistic function. A logistic link function would have the additional benefit of automatically bounding $\fesc$ to be between 0 and 1.

In principle, it may be possible to validate the method at redshifts $z\approx 0$--4, where leaking LyC flux can be directly measured (albeit with non-negligible IGM corrections at $z\approx 3$--4) and the set of emission lines we consider remain within reach of existing instruments. The prospects of carrying out such a test would however first need to be explored using simulations of $z\approx 0$--4 galaxies, since objects at these redshifts exhibit a number of properties (old underlying stellar populations, more dust attenuation and higher metallicities) that make them more complicated to analyze using the set of spectral $\fesc$ diagnostics we have so far considered. 

\acknowledgments
Erik Zackrisson acknowledges research funding from
the Swedish Research Council (project 2011-5349). Kristiaan Pelckman's research is supported in part by Swedish Research Council under contract 621-2007-6364.

\appendix
\section{Model robustness}
\label{sec:bootstrap}

The models presented in \autoref{sec:results} are only estimates in a statistical sense, since they have been derived from a finite number of samples. It is therefore desirable to investigate the statistical ``robustness'' of these estimates. If we were to train a lasso model on data outside of the original training set, would we end up with the same model, or at least a similar one? The adaptive nature of the lasso procedure makes traditional approaches in statistical inference unsuitable for answering this question, since these often rely on idealized assumptions about the distribution of the estimated variable. The bootstrap method \citep{efron1979,efron1993,hastie2015statistical} is a popular non-parametric tool for assessing the statistical properties of complex estimators (such as the lasso), when parametric inference is impossible or  a closed-form estimate of the standard errors is unavailable.

\begin{figure}[htb]
\centering
\includegraphics[width=0.9\columnwidth]{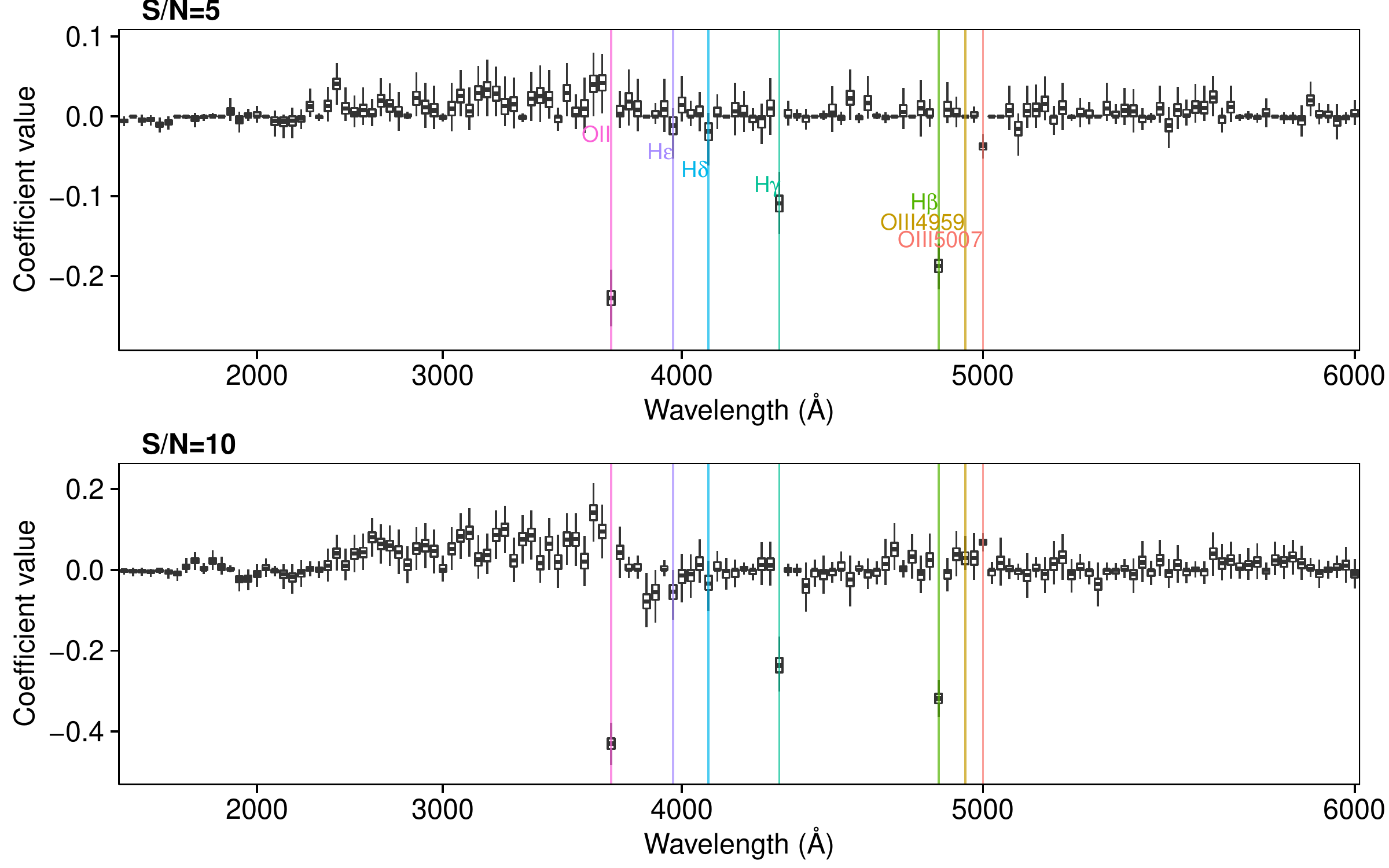}
\caption{Results of fitting lasso regression models to 1000 bootstrap samples the ShGeLMC simulation at two different noise levels. For each of the 1000 bootstrap samples, the best-fit model was found using cross-validation. The boxplots show the distribution of each individual coefficient across the samples. For clarity, the wavelength bins are placed at equal distances from each other, resulting in a non-linear $x$-axis.}
\label{fig:bootstrap_coeffs}
\end{figure}

Each bootstrap iteration is performed by randomly drawing $m$ (same size as the original data set) samples with replacement from the original data set, and selecting a model from the resampled data according to the same procedure that was used in \autoref{sec:results}. \autoref{fig:bootstrap_coeffs} shows box-and-whiskers plots of coefficients calculated for 1000 bootstrap iterations, for different signal-to-noise ratios, using two different simulations. Similar to the coefficients derived from the original data sets (see \autoref{fig:shimizu_geneva_peilmc}), the coefficients with the highest absolute values are located at wavelength bins containing prominent emission lines.

We see from \autoref{fig:bootstrap_coeffs} that the majority of the coefficients that had the highest absolute values in the models derived from the original data, consistently have the highest absolute values also in the bootstrapped models. In fact, the models trained at $S/N=5$ all look very similar to the ones shown in \autoref{fig:shimizu_geneva_peilmc}. The models change a little as the signal-to-noise ratio is increased. With less noise, fewer emission lines get selected in general. This is because the emission lines---especially those originating from the same element---are highly correlated and thus contain the same information. In noisy data, it is still advantageous to include several highly correlated lines since having multiple data points brings down the effective noise level. When the noise is lowered, this advantage is less pronounced, and the model selects only one or two lines from each element.

Based on the behavior of the cross-validated estimate of the mean-square prediction error calculated on the original data set (see  \autoref{fig:shimizu_geneva_peilmc}),
we expect to see some instability concerning the smaller coefficients in-between the strong emission lines, and at the bluer part of the spectra. That is, even if the variance of the estimated error is small, we expect that the location of the minimum error might change slightly for each bootstrap iteration, which will produce a relatively large change in the model complexity. However, as was argued in \autoref{sec:results}, the error curve indicates that adding or removing these small coefficients is not likely to have a major effect on the prediction error. Note that the cross-validation plot does not show the variance in error, and we should expect the variance to go up as the noise level on the spectra is increased. Bootstrapping is also likely to increase the variance in the error estimate. 

\bibliography{refs}

\end{document}